# A twist in chiral interaction between biological helices

A. A. Kornyshev

*Research Center "Jülich", D-52425 Jülich, Germany*

S. Leikin[*]

*Laboratory of Physical and Structural Biology, National Institute of Child Health and Human Development, National Institutes of Health, Bethesda, MD 20892, USA*

Using an exact solution for the pair interaction potential, we show that long, rigid, chiral molecules with helical surface charge patterns have a preferential interaxial angle $\sim \sqrt{RH}/L$ where $L$ is the length of the molecules, $R$ is the closest distance between their axes, and $H$ is the helical pitch. Estimates based on this formula suggest a solution for the puzzle of small interaxial angles in α-helix bundles and in cholesteric phases of DNA.

The existence of all living things depends on the molecular chirality of helices inside them. Chiral amino acids form chiral α-helices that self-assemble into structural domains of many proteins. Chiral DNA forms cholesteric liquid crystals right inside living cells [1].

Chiral interactions between biological helices present many puzzles [2]. In the cholesteric phase (Fig. 2), DNA molecules rotate by a fraction of a degree from layer to layer [1]. Their interaxial angle is much smaller than expected [3], and so is the angle between long α-helices in proteins and in membranes. This produces a macroscopic cholesteric pitch, ~0.4-5 μm in DNA [1] and even larger in cholesteric assemblies of α-helices in organic solvents [4]. One of the challenges of the physics of chiral macromolecules is to understand how the interaxial angle is encoded in intermolecular interactions [3].

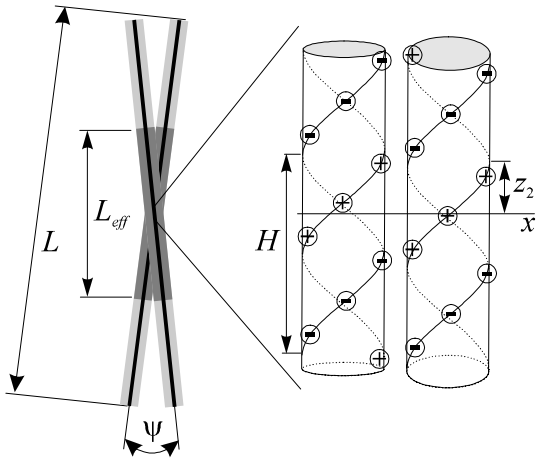

Fig. 1 A simplified, heuristic model of interaction between two long, net-neutral helical macromolecules in a nonpolar environment. *On the left*, the molecules are shown schematically by solid lines, their electric fields by light gray shading, and the overlap (interaction) area by dark gray shading. *On the right*, a small fragment of the overlap area is magnified in a side view. We assume that the molecular surface charge pattern is composed of one negatively and one positively charged, thin spiral lines shifted with respect to each other by approximately one half of the pitch. To describe the position of the strands on each helix (ν=1,2), we use the axial distance $z_\nu$ between the positively charged strand on each molecule and the x-axis connecting the points of the closest approach between molecules (as shown for molecule 2 on the right). At $z_1=z_2$, the positively charged strand of one molecule opposes the negatively charged strand of the other molecule creating a strong attraction. We assume that the molecules cross each other approximately in the middle.

In the present paper, we suggest an explanation for the small angle puzzle. We start from a heuristic model of interaction between two long, net-neutral helices in an electrolyte-free environment (Fig. 1). We present qualitative arguments followed by rigorous derivations based on an exact solution for the interaction potential. We apply the model to α-helix bundles in proteins and address the origin of the cholesteric pitch in DNA, extending our ideas to multimolecular interaction in an electrolyte solution.

*Balance of forces.* Consider two identical, right-handed, rigid, net-neutral helices (Fig. 1) immersed in an electrolyte-free medium and forming a small interaxial angle $\psi$ ($|\sin\psi|\ll 1$). The length of each molecule, $L$, is much larger than its helical pitch, $H$.

Each helix produces an electric field that decays as $\exp(-gr)$ with the distance $r$ away from it ($g=2\pi/H$) [5]. The electric fields overlap over the length $L_{eff}(\psi)$ which rapidly varies with $\psi$ (Fig. 1), e.g., $L_{eff}(\psi)$ diverges at $\psi\to 0$ for infinite helices. The energy of interaction between helices is $E_{int}=u(\psi)L_{eff}$, where $u(\psi)$ is the linear energy density. $u(\psi)$ does not diverge at $\psi\to 0$ and it can be expanded at small $\psi$ so that

$$E_{int} = \left[u_0(R,g) + u_1(R,g)\psi + ...\right]L_{eff}(\psi). \tag{1}$$

Here $R$ is the closest approach distance between the molecular axes and $u_1(R,g)\psi$ is the energy density of chiral interaction that defines the *direction* of favorable twist [6].

The helices experience two torques, the "overlap torque" ($t_1 = -u\left[dL_{eff}/d\psi\right]$) and the "chiral torque" ($t_2 = -\left[du/d\psi\right]L_{eff}$). If the helices are free to rotate around their axes, they will always select the most favorable alignment of their strands at which $u<0$ and the molecules *attract* each other (Fig. 1). Then, the overlap torque tends to reduce $|\psi|$ to maximize the attraction. On the contrary, the chiral torque tends to increase $|\psi|$. The competition between these torques establishes a nonzero equilibrium interaxial angle.

At small $\psi$, there are two distinct regions with different behavior of $L_{eff}(\psi)$:

1. ($|\psi|>\psi_*$) Tips of the helices are separated by more than $g^{-1}$. They contribute little to the interaction and $L_{eff}\propto 1/(g|\sin\psi|)$ (Fig. 1), i.e.

$$L_{eff}(\psi) = \frac{\gamma(Rg)}{g|\sin\psi|}. \tag{2}$$

The coefficient $\gamma(Rg)$ is derived below (see Eq. (10)). Within this region, $t_1/t_2 \approx \left(u_0/u_1|\psi|\right) \gg 1$ since $|\psi|\ll 1$. The overlap torque wins and it reduces $|\psi|$.

2. ($|\psi|\ll\psi_*$) Tips of the helices overlap and $L_{eff}$ stops following Eq.(2). Instead, it levels off at the value of the helix length, $L$. Since $L_{eff}$ has a maximum at $\psi=0$, $\left(dL_{eff}/d\psi\right)_{\psi\to 0}\to 0$ and $t_1/t_2\ll 1$. The chiral torque wins in this region.

The torques become equal at the crossover from the first to the second region. Exact evaluation of the equilibrium interaxial angle requires the exact $L_{eff}(\psi)$ in the crossover range. However, we can estimate that the crossover occurs at $|\psi|\sim\psi_*$, where

$$\psi_* = \frac{\gamma(Rg)}{Lg} \tag{3}$$

is obtained by extrapolating Eq. (2) to $L_{eff}(\psi)=L$. The value of $\psi_*$ may serve as an upper estimate for the interaxial angle. According to Eq. (3), this angle is small for long molecules. After deriving Eqs. (1)-(3) rigorously, we will show that this may explain small interaxial angles in *in vitro* and *in vivo* helical aggregates.

*Rigorous derivation.* The charge pattern on each molecule shown in Fig. 1 can be described in its own cylindrical coordinate frame with the long axis of the helix as the *z*-axis. The Fourier transform of the charge density in the molecular frame is given by

$$\tilde{\sigma}_\nu(q,n) = \frac{Ze_0 g}{2\pi a}(1-(-1)^n)\exp(ingz_\nu)\delta(q+ng), \qquad (4)$$

where $\tilde{\sigma}_\nu(q,n) = (2\pi)^{-1}\int_0^{2\pi}d\phi\int_{-\infty}^{\infty}dz\,\sigma_\nu(z,\phi)e^{in\phi}e^{iqz}$, $\nu(=1,2)$ labels the molecules, $e_0$ is the elementary charge, $Z$ is the number of elementary charges per helical pitch on each strand, and $z_\nu$ defines the alignment of each helix (Fig. 1).

After substitution of Eq. (4), into the expression for $E_{int}$ derived in [6], we find

$$\frac{E_{int}(\psi\neq 0)}{k_B T} = -\frac{2Z^2 l_B g}{\pi|\sin\psi|} \sum_{\substack{i,j=-\infty\\ n=2i+1\\ m=2j+1}}^{\infty} \frac{\cos[ngz_1 - mgz_2]}{|m|\sqrt{1+\tilde{w}_{n,m}^2(\psi)}} I_n(|n|ga)I_m(|m|ga)e^{-|m|Rg\sqrt{1+\tilde{w}_{n,m}^2(\psi)}} \\ \times\left(\sqrt{1+\tilde{w}_{m,n}^2(\psi)}+\tilde{w}_{m,n}(\psi)\right)^n\left(\sqrt{1+\tilde{w}_{n,m}^2(\psi)}+\tilde{w}_{n,m}(\psi)\right)^m \qquad (5)$$

where $a$ is the radius of the helices, $\varepsilon$ is the dielectric constant of the medium, $l_B = e_0^2/\varepsilon k_B T$ is the Bjerrum length, and

$$\tilde{w}_{n,m}(\psi) = \frac{ng - mg\cos\psi}{|mg|\sin\psi}. \qquad (6)$$

Eq. (5) is valid at all $\psi\neq 0$, when the molecules are long enough so that their tips do not contribute much to the interaction. At $\psi=0$ (see Ref. [6]),

$$\frac{E_{int}(\psi=0)}{k_B T} = -\frac{4Z^2}{\pi^2}L l_B g^2 \sum_{\substack{i=0\\ n=2i+1}}^{\infty}\cos[ng(z_1-z_2)][I_n(nga)]^2 K_0(nRg) \qquad (7)$$

In Eqs. (5) and (7), $I_n(x)$ and $K_0(x)$ are modified Bessel functions. At $L\to\infty$, Eqs. (5) and (7) are **exact**. They contain no approximations or assumptions, except for the choice of the surface charge distribution, Eq. (4). A similar result, can be obtained for any helical charge pattern [6].

Typically $Rg\gg 1$ [7], and the series in Eqs. (5) and (7) rapidly converge because of $\exp(-|n|Rg)$. Therefore, the summation can be truncated after $n,m=\pm 1$ so that

$$\frac{E_{\text{int}}(\psi \neq 0)}{k_B T} \approx -\frac{4Z^2 l_B g \left[I_1(ga)\right]^2}{\pi |\sin \psi|} \left\{ \frac{e^{-\frac{Rg}{|\cos(\psi/2)|}}}{|\cos(\psi/2)|} \left(1 + \frac{\sin \psi}{2|\cos(\psi/2)|}\right)^2 \cos\left[g(z_1 - z_2)\right] \right.$$
$$\left. + \frac{e^{-\frac{Rg}{|\sin(\psi/2)|}}}{|\sin(\psi/2)|} \left(1 - \frac{\sin \psi}{2|\sin(\psi/2)|}\right)^2 \cos\left[g(z_1 + z_2)\right] \right\}. \quad (8)$$

By plotting $E_{\text{int}}$, one can easily find that the energy has a minimum at $z_1 = z_2$ and $\psi \to 0$ [8].

Expanding $E_{\text{int}}(\psi \neq 0)$ in small $\psi$ at $z_1 = z_2$ and comparing the result with Eq. (7), we recover Eqs. (1), (2), where

$$\frac{u_0(R, g)}{k_B T} = \frac{u_1(R, g)}{k_B T} = -\frac{4Z^2}{\pi^2} l_B g^2 \left[I_1(ga)\right]^2 K_0(Rg), \quad (9)$$

and

$$\gamma(Rg) = \frac{\pi e^{-Rg}}{K_0(Rg)}, \quad (10)$$

After substitution of Eq. (10) into Eq. (3) and using that $Rg \gg 1$, we find

$$\psi_* \approx \frac{\sqrt{2\pi R/g}}{L} = \frac{\sqrt{RH}}{L}, \quad (11)$$

which is a remarkably simple combination of the only three length scales in the system. The energy gain upon the twist from $\psi=0$ to $\psi \sim \psi_*$ is

$$\frac{E_*}{k_B T} \sim \frac{u_1(R, g) \psi_* L}{k_B T} \approx \frac{2Z^2 l_B}{\pi^2 a} e^{-g(R - 2a)} \quad (12)$$

where we used the asymptotic behavior of $K_0(Rg)$ and $I_1(ga)$ at large $ga$.

*Interaction between α-helices in proteins.* Many proteins incorporate bundles of α-helices. The backbone of each α-helix contains a spiral of negatively charged carbonyl oxygens and a shifted spiral of positively charged amide hydrogens. In terms of the charge distribution, it resembles the heuristic model analyzed above where $a \approx 2.3$ Å, $H \approx 5.5$ Å, $L/H \approx 3\text{-}10$, $R \approx 7\text{-}12$ Å [9,10], $Z \approx 1.7$ (~$0.5 e_0$ per carbonyl or amide, ~3.5 groups per helical turn), and $\varepsilon \approx 2$ ($l_B \approx 300$ Å). For such helices, we find $\psi_* \sim 0.1\text{-}0.5$ rad (7-30 deg).

Of course, amino acid side chains impose packing constraints that affect interaxial angles and play a major role in determining the structure of α-helix bundles in proteins [9]. Still, electrostatic interaction between backbones of α-helices that do not have bulky side chains ($R \approx 7\text{-}8$ Å) is energetically significant ($E_* \sim 2\text{-}5\ k_B T$) and it may be an important player as well [11]. For instance, steric interactions define a set of preferential interaxial angles rather than a single angle [9]. Electrostatics may then determine which angle from the set is most favorable. It may not be a coincidence that the average observed angle (~19 deg [9]) is in the middle of the range predicted for chiral electrostatic interactions.

*Cholesteric pitch of DNA.* Concentrated solutions of 500-Å-long DNA fragments in 10-

300 mM salt form a cholesteric phase (see Fig. 2) at 32 Å<$R$<49 Å [12,13]. Such DNA fragments are short enough to behave as rigid rods [14] and long enough to have many helical turns, $L/H \approx 15$.

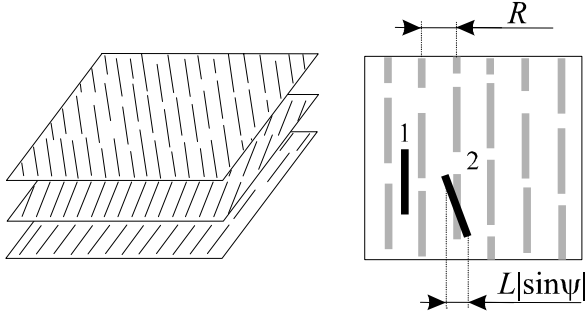

Fig. 2 Alignment of DNA helices in the cholesteric phase. *Left*: A sketch of the cholesteric phase that consists of layers of parallel molecules. Each layer is slightly rotated with respect to the layer underneath. *Right*: Top view of a molecular layer showing most favorable alignments of molecules in the layer above (black rods): (1) when molecules are homogeneously charged and (2) when molecules have helical patterns of fully balanced surface charges.

Direct measurements of intermolecular forces demonstrated that the energetics of this cholesteric phase is determined primarily by electrostatic interactions [15]. The interactions are essentially pairwise since each DNA helix overlaps with only one molecule in the layer below it [16]. Furthermore, only nearest neighbor pairs contribute to the energy because of the rapid, exponential decay of the field.

The net interaction between each two molecules can be viewed as a sum of two forces. The first one is a repulsion due to the fraction of DNA charge not balanced by bound or condensed counterions (~20-25% of "naked" DNA charge [17]). This repulsion is the same as between homogeneously charged cylinders in an electrolyte solution [6]. It favors parallel ($\psi=0$) alignment of helices in a multimolecular ensemble (Fig. 2) because this maximizes intermolecular separation and reduces the repulsion [2,18].

The second force is due to the compensated part of DNA charge. At optimal molecular alignment, it is an attraction between negatively charged phosphate strands on one molecule and positively charged grooves on the opposing molecule [5,6,19]. The physics of this force is the same as in our heuristic model (Fig. 1), but some details are different. Specifically, DNA is a double-stranded helix and it has a more complicated surface charge pattern than in Fig. 1 [20]. In addition, electrolyte reduces the decay length of the electric field from $g^{-1}$ to $\left(g^2 + \kappa_D^2\right)^{-1/2}$, where $\kappa_D^{-1}$ is the Debye length. These details are important for accurate quantitative predictions [6], but for order-of-magnitude estimates they can be neglected. From Eqs. (11), (12), where $L \approx 500$ Å, $a \approx 10$ Å, $H \approx 34$ Å, $R \sim 40$ Å, $|Z| \approx 40$ [21], and $l_B \approx 7$ Å ($\varepsilon \approx 80$), we find $\psi_* \approx 0.07$ rad and $E_* \approx 2$-$5\ k_B T$ [20].

The competition between the first force favoring $\psi=0$ and the second force favoring $|\psi| \sim \psi_*$ determines whether DNA will form a cholesteric phase and the pitch of this phase [22]. Even without a many-body statistical theory, it is clear that the equilibrium interaxial angle should be smaller than $\psi_*$ and, therefore, the expected pitch should be larger than $P_* = 2\pi R/\psi_* \approx 0.4$ μm. This is in full agreement with experimentally measured values that lie between 0.4 μm and ~5 μm [1,12,13]. Also in agreement with experiments [12,13], the pitch should not depend much on the salt concentration at 10-300 mM salt. Indeed, since $g \approx 0.2$ Å$^{-1}$, $P_* \approx \left(g^2 + \kappa_D^2\right)^{1/4} (2\pi R)^{1/2} L$ is almost unaffected by the corresponding variation of $\kappa_D$ from 0.03 to 0.2 Å$^{-1}$.

DNA molecules much longer than one persistence length also form a cholesteric phase with about the same pitch as 500 Å fragments [1]. In this case, the theory is more complicated since our description of molecules as straight, rigid rods does not apply. However, assuming that each molecule behaves as a collection of independent, one-persistence-length-long fragments [14], we can explain the observations.

Thus, the macroscopic cholesteric pitch in DNA aggregates may have the following origin. When tips of one-persistence-length long DNA fragments are separated (in lateral projection) by more than $g^{-1}$, the attraction between negatively charged phosphate strands and positively charged grooves tends to reduce $\psi$. When the tips are separated by $\geq \kappa_D^{-1}$, the repulsion associated with the uncompensated charge of DNA also tends to reduce $\psi$. These two forces overpower the chiral torque, reducing the separation of the tips and resulting in the pitch $\geq \left(g^2 + \kappa_D^2\right)^{1/4} (2\pi R)^{1/2} L \sim 8L \sim 0.4$ µm. A more detailed theory, based on the same ideas, also shows why the chiral torque may disappear at $R \leq 32$ Å (a possible solution for the puzzle of nematic-to-cholesteric transition at $R=32$ Å) and it demonstrates that the direction of the chiral torque can be reversed upon a change in counterion binding pattern or in separation [6].

*In conclusion*, let us emphasize that the goal of this letter is to illustrate on a conceptual level why chiral helical molecules may not twist more than a couple degrees in the cholesteric phase of DNA and in bundles of long α-helices. Of course, estimates are not a substitute for an accurate statistical theory that accounts for the pair potential between helices of finite length and for thermal motion. They are intended only to pave the way. The agreement with experiments indicates, however, that we may be on the right track.

We thank V.A. Parsegian and D.C. Rau for useful discussions. This work was started within the 1998 program "Electrostatic Effects in Complex Fluids and Biophysics" at the Institute for Theoretical Physics, University of California at Santa Barbara. The program was supported by the NSF Grant No. PHY94-07194. In addition, AAK acknowledges the financial support of his visits to Bethesda by the National Institute of Child Health and Human Development, NIH which allowed to complete this project.

**References and Footnotes**

DNA the cost of rotation is much larger than $k_B T$ (except at the very edge of existence of the cholesteric phase of DNA, $R \approx 50$ Å). This, in the first approximation, thermal rotations can be neglected.